\documentclass[aps,prl,twocolumn,showpacs,superscriptaddress]{revtex4}
\usepackage{amsmath}
\usepackage{dcolumn}
\usepackage{epsfig}
\usepackage{graphicx}
\usepackage{latexsym}

\begin{document}



\title{STM imaging of electronic waves on the surface of Bi$_2$Te$_3$: topologically protected surface states and hexagonal warping effects  }

\author{Zhanybek Alpichshev}
\affiliation{Stanford Institute for Materials and Energy Sciences,
SLAC National Accelerator Laboratory, 2575 Sand Hill Road, Menlo Park, California 94025}
\affiliation{Geballe Laboratory for Advanced Materials, Stanford University, Stanford, California, 94305}
\affiliation{Department of Physics, Stanford University, Stanford, CA 94305}
\author{J. G. Analytis}
\affiliation{Stanford Institute for Materials and Energy Sciences, SLAC National Accelerator Laboratory, 2575 Sand Hill Road, Menlo Park, California 94025}
\affiliation{Geballe Laboratory for Advanced Materials, Stanford University, Stanford, California, 94305}
\author{J.-H. Chu}
\affiliation{Stanford Institute for Materials and Energy Sciences, SLAC National Accelerator Laboratory, 2575 Sand Hill Road, Menlo Park, California 94025}
\affiliation{Geballe Laboratory for Advanced Materials, Stanford University, Stanford, California, 94305}
\affiliation{Department of Applied Physics, Stanford University, Stanford, CA 94305}
\author{I.R. Fisher}
\affiliation{Stanford Institute for Materials and Energy Sciences,
SLAC National Accelerator Laboratory, 2575 Sand Hill Road, Menlo Park, California 94025}
\affiliation{Geballe Laboratory for Advanced Materials, Stanford University, Stanford, California, 94305}
\affiliation{Department of Applied Physics, Stanford University, Stanford, CA 94305}
\author{Y.L.Chen}
\affiliation{Stanford Institute for Materials and Energy Sciences,
SLAC National Accelerator Laboratory, 2575 Sand Hill Road, Menlo Park, California 94025}
\affiliation{Geballe Laboratory for Advanced Materials, Stanford University, Stanford, California, 94305}
\author{Z.X. Shen}
\affiliation{Stanford Institute for Materials and Energy Sciences,
SLAC National Accelerator Laboratory, 2575 Sand Hill Road, Menlo Park, California 94025}
\affiliation{Geballe Laboratory for Advanced Materials, Stanford University, Stanford, California, 94305}
\affiliation{Department of Physics, Stanford University, Stanford, CA 94305}
\affiliation{Department of Applied Physics, Stanford University, Stanford, CA 94305}
\author{A. Fang}
\affiliation{Geballe Laboratory for Advanced Materials, Stanford University, Stanford, California, 94305}
\affiliation{Department of Applied Physics, Stanford University, Stanford, CA 94305}
\author{A. Kapitulnik}
\affiliation{Stanford Institute for Materials and Energy Sciences,
SLAC National Accelerator Laboratory, 2575 Sand Hill Road, Menlo Park, California 94025}
\affiliation{Geballe Laboratory for Advanced Materials, Stanford University, Stanford, California, 94305}
\affiliation{Department of Physics, Stanford University, Stanford, CA 94305}
\affiliation{Department of Applied Physics, Stanford University, Stanford, CA 94305}

\date{\today}

\begin{abstract}
Scanning tunneling spectroscopy studies on high-quality Bi$_2$Te$_3$ crystals exhibit perfect correspondence to ARPES data, hence enabling identification of different regimes measured in the local density of states (LDOS).  Oscillations of LDOS near a step are analyzed.  Within the main part of the surface band oscillations are strongly damped, supporting the hypothesis of topological protection. At higher energies, as the surface band becomes concave, oscillations appear which disperse with a particular wave-vector that may result from an unconventional hexagonal warping term.
\end{abstract}


\maketitle

A new type of three-dimensional (3D) bulk insulating materials with surface Quantum Spin Hall Effect states protected by time reversal symmetry has been recently predicted \cite{fu}, and soon afterwards observed experimentally in BiSb bulk crystals \cite{hsieh}. Subsequently, Bi$_2$Te$_3$ has been argued to be a similar three-dimensional topological insulator (TI), exhibiting a bulk gap and a single, non-degenerate Dirac fermion band on the surface \cite{qi1}. Indeed, recent   angle resolved photoemission spectroscopy (ARPES) confirmed that prediction \cite{chen}. In particular, with appropriate hole-doping, the Fermi level could be tuned to intersect only the surface states, indicating fully gapped bulk states as is expected from a three-dimensional TI. While ARPES could confirm the nature of the band, it is still a challenge to demonstrate unambiguously the topologically ``protected" nature of the surface state in Bi$_2$Te$_3$, or any other 3D TI system.

In this paper we present scanning tunneling microscopy (STM) and spectroscopy (STS) studies on high-quality doped Bi$_2$Te$_3$ crystals. First we show that the STS spectra exhibit remarkable correspondence to ARPES data, hence enabling us to identify each region of the local density of states (LDOS) measured. Second, by analyzing the electron-waves (Friedel-oscillations) observed near cleavage steps, we show that within the main part of the surface state band  oscillations are strongly damped, a hallmark of the strong supression of backscattering, hence supporting the hypothesis of a protected band. Finally, we show that in the region in which the surface band is warped, pronounced oscillations appear, with a distinct nesting wave-vector. Possible influence of the bulk conduction band on the oscillations is also proposed.

\begin{figure}[h]
\includegraphics[width=1.0 \columnwidth]{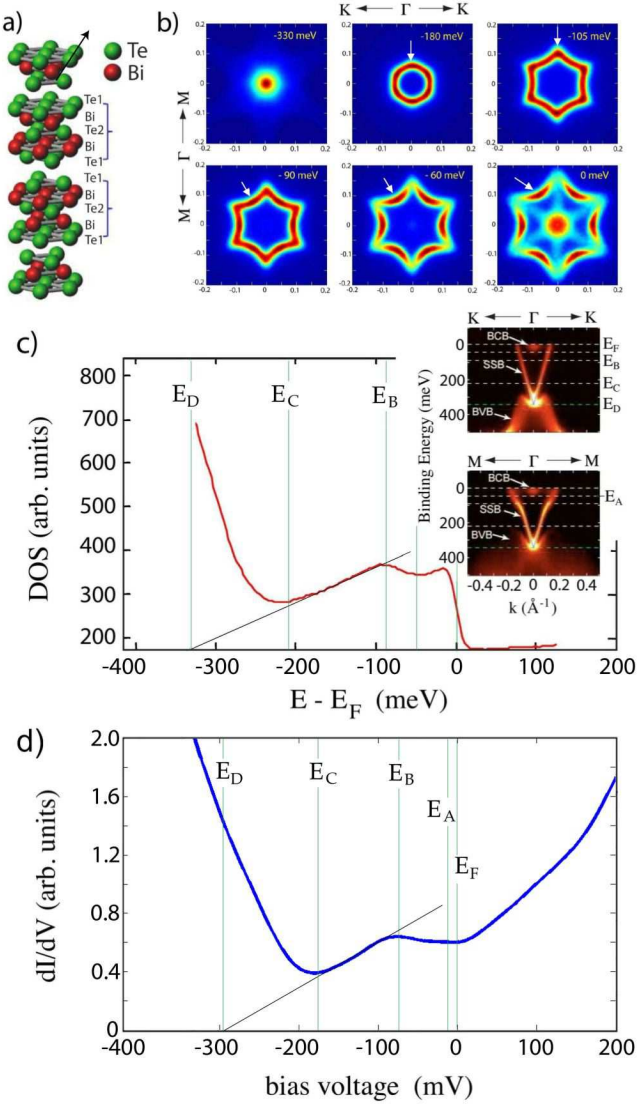}
\caption{ DOS of Sn-doped Bi$_2$Te$_3$. a) Crystal structure showing three quintuple units. b) ARPES-measured Constant-energy contours of the SSB. Strength of DOS grows from blue (no DOS) to dark red (strong contribution to DOS). Arrows point to the position of maximum in DOS (note six fold symmetry). c) Integrated DOS from ARPES. Here: $E_F$ is the Fermi level; $E_A$ is the bottom of the BCB; $E_B$ is the point of the ``opening up" of the SSB; $E_C$ is the top of the BVB; and $E_D$ is the Dirac point. d) Typical STS spectrum of a $0.27 \%$ Sn-doped Bi$_2$Te$_3$ sample similar to (c). Note the shift of all energies by $\sim 40$ meV as a result of the doping.} 
\label{stmarpes}
\end{figure}

For the present study we used Sn and Cd doped single crystals of Bi$_2$Te$_3$ (see  Fig.~\ref{stmarpes}a for crystal structure and  Fig.~\ref{stmarpes}b,c for ARPES data). Nominal doping levels between 0 and 0.27$\%$ for Sn, and up to 1$\%$ for Cd were incorporated to compensate n-type doping from vacancy and anti-site defects that are common in the Bi$_2$Te$_3$ system. Actual doping was determined separately using chemical and Hall-effect methods and were shown by ARPES \cite{chen} to be in excellent agreement with the relative position of the Dirac point with respect to the Fermi energy. For example, undoped crystals exhibit a Dirac point at $\sim -335$ meV, 0.27$\%$ Sn doping yielded a Dirac point at $\sim - 300$ meV, while for a typical $\sim 1\%$ Cd-doped crystal used for this study we observed the Dirac point at $\sim  -275$ meV, equivalent to Sn doping  of $\sim 0.45 \%$. 

Samples were cleaved in vacuum of better than $5\times 10^{-10}$ Torr, and quickly lowered to the $\sim$9 K section of the microscope, where cryo-pumping ensures that the surface remains free from adsorbates for weeks or longer.  Topography scans were taken at several bias voltages and setpoint currents (usually 200mV and 100pA).   A typical spectrum, taken at a random position on the surface of a Cd-doped cleaved crystal  is shown in Fig.~\ref{stmarpes}d. Many topography and spectroscopy maps were collected for both, the Sn-doped and Cd-doped samples with scan size as large as $\sim (360 \times 360)$\AA$^2$. Both were found to be similar to previously published STM studies of Bi$_2$Te$_3$ and Bi$_2$Se$_3$  \cite{urazhdin1,urazhdin2,yazdani}, including the typical triangular and clover-shaped defects. Weak atomic resolution could be obtained only near defects, presumably due to the overall high conductivity at the surface.

To understand the various regimes of the LDOS curves of Bi$_2$Te$_3$, we compare STS spectra to ARPES data. Fig.~\ref{stmarpes}b shows six constant-energy contours of the surface state band (SSB) of an undoped crystal. Starting from the ``Dirac point" near $E\approx-335$ meV, the contour changes from a circle to a hexagon, and becoming warped above about $E\approx -100$ meV. The inset to Fig.~\ref{stmarpes}c shows the full dispersion of the SSB in the two principal directions of the Brillouin zone (BZ), together with the portions of the bulk valence band (BVB) and the bulk conduction band (BCB) in its vicinity. Integrating the full $E({\bf k})$ data \cite{chen} over the BZ, we obtain the density of occupied states at the surface of the sample as a function of energy.  Fig.~\ref{stmarpes}c is the result of this integration in the relevant energy range that includes the Dirac band which resides in the $\sim$250 meV gap of this system. The directly measured LDOS of a similar crystal, as obtained by STM is shown in Fig.~\ref{stmarpes}d, where we could extend the data to positive bias. In both figures zero energy marks the Fermi level ($E_F$), $E_A$ corresponds to the bottom of the BCB as measured by ARPES and $E_B$ corresponds to the point where the SSB becomes warped (see Fig.~\ref{stmarpes}b). The linearly-dispersing Dirac band extends from its tip at $E_D$ to $\sim E_B$, while $E_C$ denotes the top of the BVB.  As can be seen from the ARPES dispersion in Fig.~\ref{stmarpes}c, the Dirac point is not exposed in this system, hence leaving us with only a relatively small range of pure linear LDOS expected from a Dirac dispersion.  The remarkable agreement between the DOS obtained from ARPES and  from STM suggests that no unusual tunneling matrix elements exist in this system to obscure the surface band.

\begin{figure}[h]
\includegraphics[width=1.0 \columnwidth]{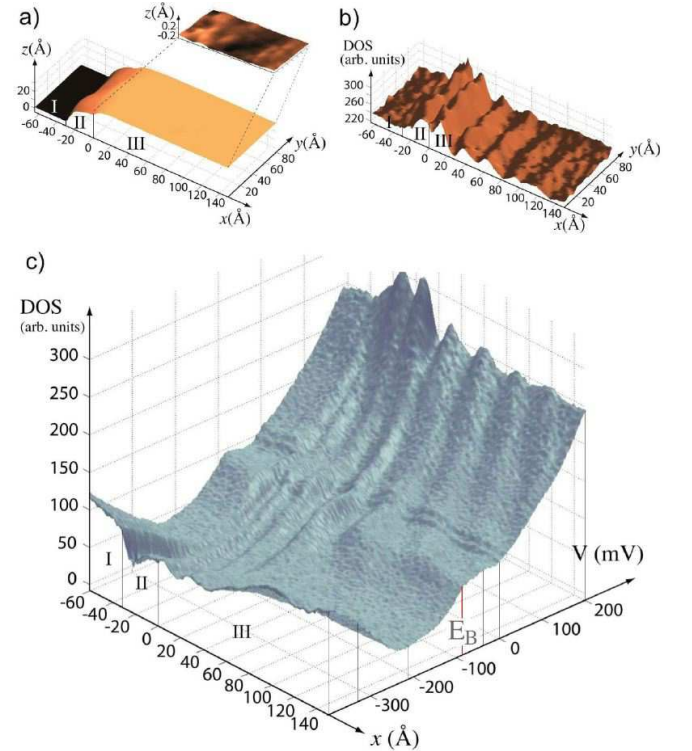}
\caption{a) Topography of the step. The total height of the step $\sim$30 $\AA$ corresponding to one unit cell. The height of region II is $\sim$20 $\AA$. Also shown is 100-fold higher resolution topograph of region III where oscillations have been studied. b) Example of (raw data) oscillations at constant energy (+200 meV) across the width of the step. c) Averaged spectroscopy as a function of distance from the step. The position of the step corresponds to ``0" of the x-axis. Note the abrupt termination of the LDOS oscillations at $E_B$. (Visible features near $E_F$ are due to convolution with the electronic spectrum of the tip arising from its sharp geometry).} 
\label{3d}
\end{figure}

As has been previously shown  \cite{crommie,hasegawa,sprunger}, the electronic properties of a surface state can be studied through the analysis of interference patterns of electron-waves (Friedel-oscillations) near defects. The simplest case is a step along a crystallographic axis, which in principle reduces the interference problem to a 1D phenomenon \cite{crommie}.  Fig.~\ref{3d}a is a topography, showing a step defect that was obtained in the process of cleaving the crystal.  The step runs along the [100] direction, thus, oscillations will be searched primarily along the [120] direction. The down-step is marked as region-I in the figure. The total measured thickness of the step between regions I and III is $\sim$30.5 $\AA$, in excellent agreement with the thickness of one unit cell \cite{bite} (i.e. three quintuples). Examination of the topography also reveals a partial cleavage area, $\sim$6$\AA$ thick, intermediate between regions I and III and we mark it as region-II.  We will concentrate on the very flat region-III, in which this combined defect is expected to scatter quasiparticles and yield interference patterns that reflect the properties of the different bands that will be probed as a function of bias voltage. We mark the beginning of the step with $x=0$ (we note that since the edge of the step is not atomically smooth, this choice is arbitrary to within a few $\AA$). We will explore the electronic structure along a straight section of the step, a strip which is 80 $\AA$ wide and 150 $\AA$ long.

Typical variation of LDOS oscillations across the step are shown for  $V=+$200 mV in Fig.~\ref{3d}b. Averaging similar patterns over the width of the step for each energy in the interval between $V=-$400 mV and $V=+$200 mV, we obtain Fig.~\ref{3d}c. Examination of this figure yields the following initial observations: i) Friedel-like oscillations \cite{friedel} that originate from the step are observed for all energies above $\sim E_B$, which is the energy above which the SSB warps; ii) the period of the oscillations in that region increases with increasing bias voltage; and iii) the ``normal" oscillation pattern becomes strongly attenuated below $\sim E_B$ with at most two visible peaks.

\begin{figure}[h]
\includegraphics[width=1.0 \columnwidth]{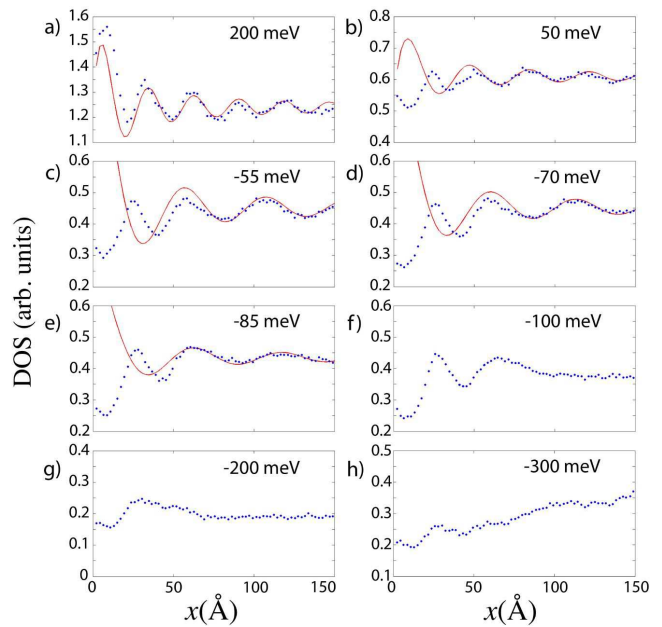}
\caption{Sample averaged LDOS  as a function of distance from the step for 8 energies.  Red lines are fits to the data (see text).  a), b), c), d), and e) correspond to energies above $E_B$ showing well pronounced  oscillations.  f), g), and h) correspond to energies within the SSB, for which ``h" may also reflect substantial influence of the BVB.}
\label{osc}
\end{figure}

We now turn to the analysis of the data. Fig.~\ref{osc} shows a series of 8 energies ranging from high bias that includes the BCB, to near the Dirac point (which include BVB). Panels a$-$e are in the regime of clear dispersive oscillations. Here we fit the curves to a simple  $\rho(x) \propto A(E){\rm sin}(2kx+\phi)/x$ expression, appropriate for 1D oscillations \cite{crommie}  of a surface band (with an amplitude $A$, and a scattering potential phase-shift  $\phi$). Attempts to fit the data with  decays that are either much stronger (e.g. $x^{-2}$) or much weaker (e.g. $x^{-0.5}$) gave markedly poorer fits.  In addition, pure bulk  oscillations would require an unobserved $k$-divergence at $E_A$. Thus, the data strongly suggest that all the observed oscillations originate from the SSB.  While the oscillations seem to fit well the theoretical curve away from the step, the fit near the step is poor, especially close to $x=0$. This is expected due to the step's height variation and its roughness   \cite{crommie} (Fig.~\ref{3d}a). In addition, the fit near the step may be disrupted due to a nondispersing peak at $\sim$25 meV observed from below $\sim$100 meV, down to  the Dirac point. 

\begin{figure}[h]
\includegraphics[width=1.0 \columnwidth]{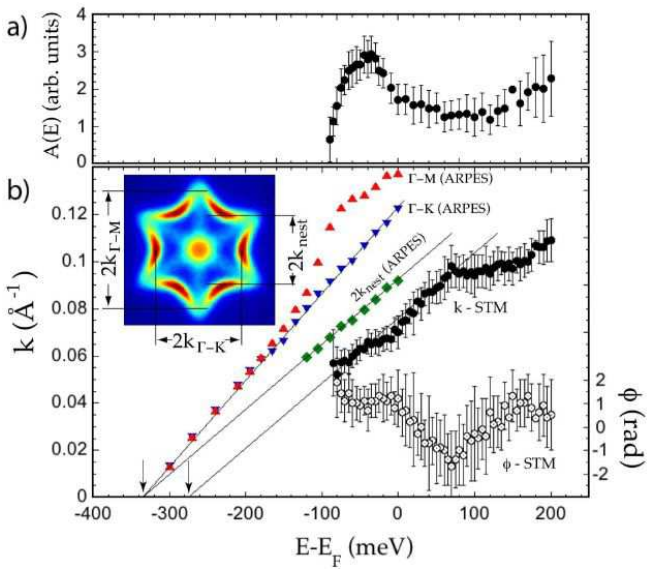}
\caption{Results of fits of oscillations (see text for details) for a Cd-doped sample with Dirac point at $\sim-275$ meV. a) Variation of oscillations amplitude. b) Oscillation wave-vectors (full circles) and phase shifts (open circles).  Triangles follow $k_{\Gamma-M}$ and  $k_{\Gamma-K}$, and diamonds follow $k_{nest}$, all three are extracted from ARPES data of an undoped sample (see inset for example).  Arrows point to the respective Dirac points.}
\label{dispers}
\end{figure}

The amplitudes of the fitted oscillations are plotted in Fig.~\ref{dispers}a, and the respective values of $k$ and $\phi$ are plotted in Fig.~\ref{dispers}b.  Also shown in Fig.~\ref{dispers}b are the  ARPES extracted dispersion of the SSB for an undoped crystal in the $\Gamma$-M and $\Gamma$-K directions. In the inset of  Fig.~\ref{dispers}b we use a $-15$ meV contour of the SSB to mark these two wavevectors as well as a ``new" wavevector, $k_{nest}$ which emerges above the energy in which the Fermi surface changes character from convex to concave. The dispersion of $k_{nest}$ agrees remarkably well with the dispersion extracted from the STM data below 70 meV, once the Dirac points are shifted from -335 meV in the undoped crystal to $\sim -275$ meV appropriate for the Cd-doped crystal.  The associated phase changes by an amount of  $\sim \pi$ in that regime. Above 70 meV a break in the  linear dispersion is observed, accompanied by a change of the phase back towards its initial value.  We believe that at this energy the BCB influences the scattering, though the oscillations are still 2D in nature as we do not observe a change in the power of the decay.
 
The abruptness with which the oscillations cease below $\sim -100$ meV (i.e. below $\sim E_B$) strongly support the prediction that Bi$_2$Te$_3$ is a TI \cite{qi1}. For any ``normal" surface state band we would expect $\sim x^{-1}$-decaying oscillations to persist throughout the band.  By contrast,   SSB in TI systems should be protected from backscattering due to their definite chirality, and thus should not exhibit oscillations, which are a consequence of the interference of incoming and backscattered waves. This ``protection" results from the definite chirality of the surface state. Thus, for scatterers that respect time reversal symmetry, any backscattering process should also include spin-flip in order to stay within the same chirality class. Since no such events are possible for non-magnetic defects, such as the step, no backscattering is allowed. However as is evident from Figs.~\ref{osc}f, g and h,  the LDOS profile in the pure Dirac regime exhibits a nondispersing peak at $\sim 25\AA$ and weak structure beyond it, suggesting at most, fast-decaying oscillations below $E_B$ and possibly a bound state near the step \cite{nondisperse}.  The idea of a bound state is also supported by the fact that the phase of the scattering potential changes by an order of $\sim \pi$ within the SSB. 

While the above analysis of the dispersion strongly  suggests that the oscillations originate  exclusively from the SSB, the BCB can still play a role in promoting oscillations if we assume that due to its presence, quasiparticles can be scattered from {\bf k} to {\bf -k} via intermediate states in BCB.  While calculations of such an effect are cumbersome and thus beyond the scope of this present work, it is not impossible that such processes maintain the $x^{-1}$ type oscillations \cite{qin}. Since processes involving  BCB are expected to affect the dispersion of the oscillations, it is reasonable to conclude that above $\sim 70$ meV the observed change in dispersion is due to BCB effects ( Fig.~\ref{dispers}b).

Maybe the most surprising observation is the emergence of linearly dispersing oscillations between $\sim -80$ meV  and  $\sim 70$ meV within the surface-state Dirac band. Here we propose that this is a direct consequence of the warping of the SSB, as its contour changes  character from a fully convex shape (smoothly evolving from the Dirac point to a circle, and to a hexagon higher up in the band) to a concave hexagram (Fig.~\ref{stmarpes}b). In addition, above $\sim E_B$ DOS shift from the tips of the hexagons to the warped sides of the hexagram allowing for a new, unprotected,  nesting wave-vector to appear. The inset in Fig.~\ref{dispers}b depicts this nesting wave-vector in the ARPES data. The linear dispersion extracted from the STM data, in perfect correspondence to $k_{nest}$ extracted from ARPES, suggest that this is the only possible wave-vector for oscillations within the warped SSB. This is easily understood if we assign a small $z$-direction component for the spin ($S_z$). Such a component will have to alternate in sign between warped hexagonal sides, hence the only possible nesting wave-vector that respects $S_z$ is the observed $k_{nest}$. Thus, the above analysis strongly suggests that the oscillations that emerge from the otherwise ``protected" SSB are a result of an unconventional hexagonal warping term  \cite{fu2} of the Bi$_2$Te$_3$ TI - system .

{\it Note added:}  Following posting of the first version of this paper on the archive, Liang Fu posted a paper \cite{fu2}  explaining the correlation between LDOS oscillations and convexity of constant energy contours. The revised version of this paper emphasizes this observation.

\acknowledgments
Discussions with Qin Liu, Xiaoliang Qi, Shoucheng Zhang, and especially Liang Fu are greatly appreciated.  This work was supported by the Department of Energy Grant  DE-AC02-76SF00515. ZA is supported by Stanford Graduate Fellowship.


\begin{thebibliography}{99}
\bibitem{nondisperse}
More work is needed to substantiate the claim that this peak is associated with a surface bound state. We find similar features near macroscopic defects in Bi$_2$Se$_3$.
\end{thebibliography}
\end{document}